\documentclass{article}

\usepackage{arxiv}

\usepackage[utf8]{inputenc} 
\usepackage[T1]{fontenc}    
\usepackage{hyperref}       
\usepackage{url}            
\usepackage{booktabs}       
\usepackage{amsfonts}       
\usepackage{microtype}      
\usepackage{graphicx}

\usepackage[numbers]{natbib}
\usepackage{doi}
\usepackage{xspace}
\usepackage[most]{tcolorbox}
\usepackage{graphicx}
\usepackage{wasysym}

\usepackage{multirow}
\usepackage[normalem]{ulem}
\useunder{\uline}{\ul}{}

\renewcommand{\sp}[0]{security and privacy\xspace}
\newcommand{\SP}{security and privacy\xspace}
\renewcommand{\paragraph}[1]{
\vspace{2mm}
\noindent\textbf{#1}
}
\newcommand{\etal}{\emph{et al.}\xspace}

\usepackage[noorphans,vskip=0.0pt,leftmargin=.1pt]{quoting}

\newtcolorbox{SummaryBox}[1]{enhanced,arc=1mm,outer arc=1mm,
  boxrule=0mm,toprule=0mm,bottomrule=0mm,left=1mm,right=1mm,leftrule=2pt,
  titlerule=0mm,toptitle=0mm,bottomtitle=0mm,top=0mm,
  colframe=blue!50!black,colback=blue!5!white,coltitle=blue!50!black,
  colbacktitle=yellow!50!white,colback=green!5!white,
  title=#1,
  fonttitle=\bfseries\sffamily\normalsize,fontupper=\normalsize\itshape,
}

\title{Investigating the Security \& Privacy Risks from Unsanctioned Technology Use by Educators}

\author{ 
\textbf{Easton Kelso}\\
\texttt{eakelso@asu.edu}\\
Arizona State University
\and
\textbf{Ananta Soneji}\\
\texttt{asoneji@asu.edu}\\
Arizona State University
\and 
\textbf{Syed Zami-Ul-Haque Navid}\\
\texttt{snavid2@asu.edu}\\
Arizona State University
\and \\
\textbf{Yan Soshitaishvili}\\
\texttt{yans@asu.edu}\\
Arizona State University
\and\\
\textbf{Sazzadur Rahaman}\\
\texttt{sazz@cs.arizona.edu}\\
University of Arizona\\
\and\\
\textbf{Rakibul Hasan}\\
\texttt{rakibul.hasan@asu.edu}\\
Arizona State University
}

\renewcommand{\shorttitle}{Education Technology Acquisition Practices}

\hypersetup{
pdftitle={Investigating the Security \& Privacy Risks from Unsanctioned Technology Use by Educators},
pdfauthor={Easton Kelso, Ananta Soneji, Rakibul Hasan},
pdfkeywords={First keyword, Second keyword, More},
}

\begin{document}
\maketitle

\begin{abstract}
Educational technologies are revolutionizing how educational institutions operate. Consequently, it makes them a lucrative target for breach and abuse as they often serve as centralized hubs for diverse types of sensitive data, from academic records to health information. Existing studies looked into how existing stakeholders perceive the security and privacy risks of educational technologies and how those risks are affecting institutional policies for acquiring new technologies. However, outside of institutional vetting and approval, there is a pervasive practice of using applications and devices acquired personally. It is unclear how these applications and devices affect the dynamics of the overall institutional ecosystem.

This study aims to address this gap by understanding why instructors use unsanctioned applications, how instructors perceive the associated risks, and how it affects institutional security and privacy postures. We designed and conducted an online survey-based study targeting instructors and administrators from K-12 and higher education institutions.
\end{abstract}


\section{Introduction}

Technology use in the education domain, at both K12 and higher education institutes (HEI), has seen unprecedented growth recently, digitizing every aspect of teaching, learning, research, and administrative tasks~\cite{datafication-HE, constant-expanding-classroom, edtech-ccs2024, chanensonUncoveringPrivacySecurityK12CHI23}. 
Simultaneously, there has been much effort from the research community to understand \SP risks as perceived by different stakeholders in this setting. For example, researchers investigated the factors affecting decisions to adopt tools by educators and administrators~\cite{chanensonUncoveringPrivacySecurityK12CHI23, balashEducatorsPerspectives2023, shiojiItsBeenLovely}, rising \SP concerns of data subjects due to an increasing number of tools being deployed~\cite{edtechPETS23, studentPrivacySchoolDevices, examining-examiners, yang_discovering_2024, radway_investigation_2024}, how these tools are being audited and maintained to alleviate \SP risks~\cite{edtech-ccs2024, chanensonUncoveringPrivacySecurityK12CHI23}, as well as how institutional policies and other regulatory measures aim to minimize the use of collected data in privacy-invasive ways~\cite{radway_investigation_2024, edtech-ccs2024}. 

Existing literature lacks an understanding of the \SP impacts of tools and services that escape institutional auditing and are nrot bound by institutional policy or contracts that mandate certain levels of data protection and restrict the use and sharing of data. A previous study noted that such use cases may be ubiquitous as there is a plethora of apps and services that can be used by anyone for teaching and learning-related activities, often free of cost~\cite{edtech-ccs2024}. Even paid apps in many cases can be personally acquired, for example using research grants or departmental funds, where they do not go through the institutional procurement and security audit process if the price is lower than a threshold~\cite{edtech-ccs2024}, and thus they do not have a formal contract restricting data collection and use. This lack of a vetting process and contracts may lead to increased risks of leaking private data, as well as legal liabilities for the users or the institutions since many institutional records are protected under laws (such as FERPA~\cite{ferpa}) that do not apply to other domains. Thus, investigating the use of technologies (apps and services) in institutional settings that were not institutionally acquired or sanctioned, and the associated data \SP issues need urgent attention.

%
This paper contributes to shedding light on this matter; specifically, it seeks to answer the following research questions: (1)~RQ1: What unsanctioned technologies do instructors (at K12 or HEIs) use that are not institutionally acquired (i.e., unsanctioned) and why?, (2)~RQ2: How do instructors perceive and experience \SP issues and risks of those apps and how those perceptions impact their use?, and (3)~How the use of unsanctioned technologies impact the \SP posture of education institutes?

To answer these questions, we first conduct an online study (N=432) involving instructors at K12 and HEIs to learn about their use of unsanctioned apps and personal devices, their perception and experience of associated \SP issues, their knowledge and understanding of institutional policy about unsanctioned technology use, as well as their efforts to minimize \SP risks. We identified 452 unique apps they use for various purposes, including research purposes and photography classes. We identify the absence of institution-provided alternatives, habituation, usability, and surprisingly, being `forced' by school admins, among the primary reasons for unsanctioned app use. Major selection criteria they used include the apps' capacity to engage students and AI (Artificial Intelligence) features; \SP were rarely considered as a primary deciding factor. 
We also find that many participants continued to use apps despite distrusting them and even after observing privacy-invasive behaviors. Less than half of the participants knew the existence of institutional policy, and most of those who knew went against it to use unsanctioned apps. 




We supplement these findings with another study surveying school administrators, IT support staff, and technology policy makers (N=29). The results concur with findings from the first survey, where admins acknowledged that instructors often use unsanctioned apps and request their integration with institutional apps, which are sometimes accommodated. Participants also listed \SP incidents their institution faced because of unsanctioned app and device use; examples include a third-party app scraping institutional data, and increased security vulnerability due to instructors forwarding emails to their private accounts. 

Overall, our studies surface striking \SP issues, which might impact millions of students, arising from using unsanctioned technology. We discuss the privacy, security, and compliance implications of these results, and provide recommendations to improve this situation. 


\section{Related work}

Technologies now manage nearly every aspect of academic activities~\cite{edtechPETS23, constant-expanding-classroom}, with tech ecosystems constantly evolving with the availability of numerous mobile apps and extensions for other platforms (such as through marketplaces for Zoom \cite{zoom} and Canvas \cite{Canvas}). 
This continuous digitization has led to security and privacy vulnerabilities: in 2023, data breaches at HEIs have cost an average of $3M$ USD~\cite{breach-stat}. 

Past research has studied the institutional use of technologies. Radway~\etal investigated if and how universities conform to the Family Education Rights and Privacy Act (FERPA) while sharing directory information~\cite{radway_investigation_2024}. Chanenson~\etal interviewed K-12 school officials and IT personnel to understand districts' use of technologies and how they manage student privacy and security~\cite{chanensonUncoveringPrivacySecurityK12CHI23}. Balash~\etal surveyed university instructors to understand the prevalence of using online exam proctoring apps and why they are (not) adopted~\cite{balashEducatorsPerspectives2023}, while Shioji~\etal investigated the same with a target audience of senior administrators~\cite{shiojiItsBeenLovely}. Kelso~\etal investigated how universities procure technology, their auditing process, and how they maintain institutional security posture. Paris~\etal reported how loopholes in regulations and institutional contracts can be exploited to invade privacy. However, the literature lacks studies on unsanctioned technology use and its impact on \SP, which is crucial for a more comprehensive understanding of educational institutes' \SP posture.
\section{Study 1: Instructor Survey}

To understand the use of unsanctioned apps and devices in K-12 and university settings, we conducted a survey in November and December 2024 targeting educators who rely on personally acquired technologies for teaching-related activities. The following sections detail the methodologies and findings of this study.

\subsection{Methods}

\paragraph{Survey design.}
We designed an online survey to capture the use of unsanctioned apps for teaching and grading activities. 
This survey included a mix of open-ended and multiple-choice questions, structured to gather comprehensive insights into app usage, institutional awareness, and data management practices.

After obtaining consent, participants were first asked to list at least three apps (for mobile, web browsers, or desktop computers) that they have used for teaching and grading-related activities and why.
Follow-up questions explored participants’ perceptions and experiences with their listed apps and their knowledge of institutional policies regarding unsanctioned app use.
The survey concluded by asking participants about the data management practices of discontinued apps, specifically whether these apps provided the option to delete user data and if participants had utilized this feature 
(full questionnaire in \S~\ref{app:study1ques}).


\paragraph{Participant recruitment.}
The survey was conducted using the Qualtrics \cite{Qualtrics} platform. We recruited participants through Prolific, a well-known platform for recruiting participants from a diverse pool. 
The study received Institutional Review Board (IRB) exempt status, as it posed minimal risk to participants. 

Selection criteria ensured that participants were current or former educators teaching at K-12 schools, community colleges, colleges, or universities, with prior experience using unsanctioned apps for teaching, grading, or related activities. 
Each participant was compensated $\$5$ for completing the survey. Our survey was designed and checked to take approximately 10 minutes. 

\subsection{Results}

\subsubsection{Participants}
We collected data from $450$ participants and manually reviewed it to exclude $18$ responses that were either autogenerated or gibberish. This left us with $432$ valid responses: $283$ from K-12 educators and $149$ from HEI educators.
Our participants represented diverse demographics: $313$ identified as female, $137$ as male, and $5$ as non-binary. 
The age distribution was as follows: $98$ participants were under $30$ (18–29), $269$ were 30–50, and $89$ were over $50$.
Regarding education, $18$ had a high school diploma, $55$ had a Bachelor’s degree, $104$ had a Master’s degree, $23$ had a doctorate, and the rest preferred not to answer.
Discipline-wise, $28$ participants taught STEM subjects, $290$ taught non-STEM subjects, and $135$ did not disclose their discipline.

\subsubsection{Use of unsanctioned apps}
\label{use_unsanctioned_apps}
Participants listed a total of 1,654 apps and services, with 494 being unique. Among these, participants K-12 and HEIs shared 88 applications. K-12 participants 284 unique apps, while HEIs identified 121 unique applications.
The most popular personal use applications were Kahoot! (n=73), ChatGPT (n=69), Google Classroom (n=63), Canva (n=55), and Quizlet (n=50). For K-12 participants alone, top 3 were Google Classroom, Kahoot!, and a tie between ChatGPT and Canva, while HEI participants favored ChatGPT, Kahoot!, and Canva.

A huge majority ($87\%$ of $432$ participants) stated that they use personal devices for teaching-related tasks.
Among these, $114$ participants used personal devices daily, while 139 used them several times a week.
Moreover, $277$ participants downloaded institutional documents (e.g., student data, grade books) onto personal devices.
Of these, $137$ reported that their devices had automatic cloud backups through personal accounts, potentially exposing institutional data to security risks. 

\subsubsection{Primary factors in app selection}
\label{app_selection_factors}
Gamification was a key factor in app selection, with $30.4\%$ of K-12 apps focused on helping students
learn basics of reading, mathematics, and other general subjects.
Classroom management tools also played a significant role among K-12 participants, comprising $12.2\%$ of the apps.
Participants emphasized their value, describing them as ``a fun way to track classroom behavior'' (P59) and noting they were ``chosen for their ability to enhance classroom management and communication with both students and parents'' (P83).

HEI educators used a range of tools, with $30.3\%$ of their apps 
focused on tasks such as note-taking, sharing materials with students, and conducting research. 
AI-based applications made up $7.4\%$ of their apps, reflecting the growing integration of AI in education. 
Educators stated that the use of AI ``makes students want to learn, and motivates me'' (P211) and ``enhances my teaching of literature'' (P210).
In addition, specialized tools were commonly used, and one participant mentioned the need for apps to ``shoot RAW images with a smartphone'' (P151).

\subsubsection{Reasons behind unsanctioned app use}
\label{reasons_unsanctioned_app_use}
In both K-12 and HEI contexts, the overwhelming majority ($n=387$) mentioned ``ease of use'' or similar phrases as the primary reason to use unsanctioned apps, even when institutional alternatives were available. 
Accessibility ($n=281$), student engagement ($n=279$), and price ($n=212$) followed closely as significant factors. 
Almost $70\%$ of the participants who prioritized engagement ($n=279$) were K-12 educators ($n=195$), emphasizing that these apps ``keep students very engaged'' (P30).
Familiarity with certain tools also played a role: $84$ participants resisted school-provided options, noting they were accustomed to other tools due to previous careers, long-term use, and unwillingness to adapt to a new tool.


Within HEI setting, primary driver was the need for specialized tools unavailable through their institutions (n=61). 
Participants highlighted requirements for tasks such as ``basic image editing'' (P300) or accessing ``reference apps for specific films and developer combinations'' (P151). 
Research-related tools were also important, with educators seeking apps to ``help with research papers'' (P52) and tools for lab-specific needs, such as ``coding observational data'' (P37).
%
%
Surprisingly, only $7/432$ (i.e., $1.6\%$) participants cited ``security'' as a factor in their unsanctioned application choices, indicating that most respondents did not prioritize it as a primary selection criterion.

\subsubsection{Security and Privacy Perceptions}
\label{sp_perceptions}
As mentioned previously, while \sp were rarely top priorities in app selection, $218$ participants acknowledged considering these issues for at least one unsanctioned app they listed, and $72$ consistently evaluated \sp for all three apps they listed. On the other hand, a significant proportion---more than one-third ($n=176$)---did not consider these aspects at all when choosing their apps. 
This highlights a complex relationship between perceived risks and actual behavior.


Despite limited consideration during selection, concerns about \sp issues emerged post-usage. 
$23$ participants reported experiencing security or privacy breaches with at least one app, while $38$ expressed concerns about apps' data collection practices potentially violating user privacy. 
Alarmingly, $43$ participants believed that at least one app sold user data to advertisers, with one participant extending this belief to all three apps they listed.

Compliance awareness was mixed. 
All participants assumed FERPA~\cite{ferpa} and HIPAA~\cite{hipaa} applied to the apps they used, but their perceptions of compliance varied.
For FERPA, 107 participants believed all three apps were compliant, 7 believed none were, and 237 thought at least one app was compliant. However, 71 participants were unsure of any app's FERPA compliance. For HIPAA, 81 participants believed all three apps were compliant, 10 believed none were, 182 believed at least one app was compliant, and 51 were unsure.

Trust in developers’ ability to safeguard user privacy was similarly divided. Fifty-three participants distrusted at least one app, and one distrusted all apps they used. Forty-six doubted the competence of at least one app’s developers to protect privacy, with one participant holding this belief for all their apps.


\subsubsection{Institutional policy about unsanctioned app use}
\label{inst_policy_on_unsanction_app_use}

Surprisingly, only $30.3\%$ of K-12 participants and $24.8\%$ HEI participants were aware that their institution had a policy regarding the use of unsanctioned education technologies. 
Many of those aware of policies (N=22) admitted to using unsanctioned apps despite policy prohibitions, citing institutional shortcomings.
P10 shared that ``[my institution doesn’t] provide a workable path to utilize new and emerging technologies so [I] have to go against it.'' 
Another noted, ``we are not supposed to use any application that is [not] already approved, but many, like myself, ignore the rule.'' (P68).

$107$ participants reported receiving institutional warnings about the risks associated with unsanctioned app use. 
Among them, $33$ mentioned that those warnings led to behavioral changes.
The most common changes included discontinuing unsanctioned app use, switching to alternatives perceived as more secure, adopting 2FA, and exercising greater caution when sharing content with platforms. 


\subsubsection{Discontinuation and data deletion}
\label{discontinue_data_deletion}
$368$ participants reported that they continue to use the three unsanctioned applications they mentioned in the survey. 
This includes 19 participants who experienced security or privacy issues with the apps, 33 who believed the apps collect data that violates users' privacy, 33 who stated the apps share data with advertisers, 47 who expressed distrust in the apps, and 41 who considered the app providers incompetent at protecting privacy.
Eighty-five participants mentioned that they stopped using at least one app.
Only nine of them said that the apps provided an option to delete data, and eight of them requested data deletion, while 60 participants were uncertain about the data deletion feature.

\section{Study 2: Admin survey}
To understand if and how the use of unsanctioned apps by educator leads to any \SP incidents, we conducted another online study targeting administrators, IT personnel, data governance bodies, and technology policymakers from US-based educational institutes.

\subsection{Methods}

\paragraph{Survey design.}
Survey questions were created based on results from the first study. We first asked participants if they recommended unsanctioned apps to instructors, or received requests from instructors to integrate such tools with other institutional tools. Next, we asked them to explain if there was any \SP incident at the institution due to instructors using unsanctioned apps (see \S~\ref{app:study2ques}) for the full questionnaire.


\paragraph{Participant recruitment.} Large-scale recruitment of participants from the target population is infeasible~\cite{edtech-ccs2024} and there is no direct way to reach them through online survey platforms like Prolific. Thus, we attempted to recruit participants through direct emails and posting the study link to Educause platform \cite{Educause} and two Reddit 
\cite{Reddit} groups: \textit{k12sysadmin} and \textit{k12cybersecurity}. 


\subsection{Results}
\subsubsection{Participants}
We collected data from 29 participants, denonated with an 'A'. Among those, four were from K-12 institutions while the rest were from higher education institutions. Of those participants, 12 were actively in roles related to information security while the rest were in administrative roles at their institutions such as business manager, principal, or vice chancellor.

\subsubsection{Use and integration of unsanctioned apps}
Seven participants said that they had recommended unsanctioned apps to instructors, supporting results from Study 1. Their recommendations come from needing ``to meet niche demands that our institutional platforms cannot manage, or where instructors are looking for free alternatives for their use or their student's use.'' (A4). The applications administrators recommend are ``common cloud platforms'' and platforms that the wider population is using like `` Canva, Slack, and ChatGPT'' (A7).

On the other hand, all but three participants had received instructors' requests for unsanctioned tools to be integrated into institutionally licensed tools. 
For example, A4 mentioned how they ``have instructors wishing to integrate many tools such as polling, scheduling, citation software for use in classes.''.  One administrator discussed how ``Almost at an individual level, everyone has their "solution" to teaching/learning "better" and brings it into the classroom with disregard.'' (A21) regarding how educators bring unsanctioned technology in classrooms. 

Regarding fulfilling the integration requests, two participants said such requests are never accepted, others' responses varied from `sometimes' to `most of the time'. Twelve participants said such integrations go through IT audit, others were unsure about this.

\subsubsection{Security and privacy incidents}
\label{admin_sp_incidents}
Nine participants indicated they had experienced \SP incidents due to unsanctioned app use. A4 stated that they ``have had third-party integrations scrape user data without being vetted and without contractual controls in place.'' Another participant (A12) stated challenges they faced in ensuring institutional security because ``a significant subset of faculty were in the habit of auto-forwarding internal university email to personal email accounts. This was a problem for many reasons, the biggest being none of our phishing controls or detection would detect anything after such emails had transited outside our managed environment, including phishing attack successes and account compromise [\dots]'' Without providing details, one participant (A19) mentioned an issue with \textit{otter.ai},\cite{OtterAI} that was integrated with \textit{Zoom} video conferencing tool.

Several participants expressed worries about the impact unsanctioned apps can have on institutional security posture and legal compliance. Overall their top concern for the use of those technologies stems from the possibility of leakage of PII, ``Personal information collected and stored incorrectly.'' (A10). One participant explained how using unsanctioned technology ``strips the controls [the institution] very intentionally design, build in and manage'' (A11).  A21 stated that ``In the case of a classroom, [disclosure of information] can commonly meet the criteria for a FERPA breach exposing the institution to serious liabilities''.
Another participant pointed out how ``Instructors [may] not understand that the institution has laws that it must follow to protect its data… And sometimes integrations impact a security plan/posture and we just can't use them'' (A16). Concerns were also growing regarding the use of AI tools like ChatGPT. A20 mentioned how they ``believe faculty are regularly using "free" AI tools with student data'' which could lead to student data being leaked and put into those AI systems.





\section{Discussion}
This study sheds light on the nuanced challenges associated with educators' use of unsanctioned applications in educational settings. 
From \sp perceptions to institutional policy awareness and administrative roles, our findings highlight a complex interplay of factors that shape app usage behavior. 
Below, we discuss key themes and their implications.

\subsection{Educators' Security and Privacy Awareness vs. Action}
The findings reveal a striking paradox: while many educators acknowledge the importance of \sp (\S~\ref{sp_perceptions}), these concerns rarely influence app selection or usage behavior. 
For example, 218 participants considered \sp for at least one app, yet more than a third ($n=176$) did not factor these considerations into their decisions at all. Furthermore, even after experiencing breaches ($n=23$), suspecting apps of poor data collection practices ($n=38$), or believing apps sold user data ($n=43$), many educators continued to use these apps. 
This behavior underscores a tendency to prioritize ease of use, engagement, and functionality (\S~\ref{reasons_unsanctioned_app_use}) over compliance and safety, highlighting a gap between perceived risks and actionable responses.  

Compounding this issue is the widespread uncertainty surrounding regulatory compliance. Despite most participants assuming FERPA and HIPAA applied to the apps they used, many were unsure of their compliance status ($n=71$ for FERPA, $n=51$ for HIPAA). 
This lack of clarity not only compromises institutional data security but also demonstrates the inadequacy of current measures to educate and empower educators in \emph{evaluating app security}. 
Institutions must focus on bridging this gap by offering secure alternatives that align with educators' needs while providing clear guidance and training on security and compliance.

\subsection{Limited Effectiveness of Institutional Warnings}
Only $24.8\%$ of participants ($n=107$) were aware of institutional warnings about the risks of unsanctioned app use, and among those, only 33 were influenced to change their behavior (\S~\ref{inst_policy_on_unsanction_app_use}). 
These numbers suggest that current warning mechanisms fail to translate awareness into meaningful action. 
Most educators, even when aware of institutional guidelines, prioritize practicality and familiarity over compliance.  

This highlights the need for institutions to rethink their policy enforcement strategies. 
Instead of relying solely on warnings, institutions should adopt a proactive approach by educating educators on the risks associated with unsanctioned apps and offering clear, actionable alternatives. 
For example, providing training sessions on how to evaluate apps for compliance, encouraging the adoption of institution-approved tools, and implementing educator-friendly policies that balance security with flexibility could enhance adherence. Ensuring that warnings are both relatable and actionable will likely improve their effectiveness in fostering behavior change.

\subsection{The Illusion of Control}
While the ability to download unsanctioned apps may give educators a sense of flexibility and autonomy, it also creates an illusion of control, particularly regarding data security and deletion. 
Our study results highlight a significant gap in educators' understanding of app exit strategies (\S~\ref{discontinue_data_deletion}). 
Of the 85 participants who stopped using at least one app, only nine were aware of a data deletion option, and just eight requested data deletion. Meanwhile, 60 participants were uncertain about whether such options existed.  

This lack of awareness demonstrates that educators are not currently equipped to safeguard institutional data effectively, especially when using unsanctioned apps. 
Without clear exit pathways, sensitive data may remain exposed to risks long after app use ends. 
This emphasizes the critical need for institutions to implement robust policies and educational initiatives that address not only app usage but also secure disengagement practices. 
Training educators on topics such as data retention, deletion policies, and privacy risks can help mitigate these vulnerabilities while reinforcing the importance of compliance.

\subsection{Administration Challenges and Contradictions}
Administrators are often on the frontlines of managing the consequences of unsanctioned app use, including \sp incidents. 
As evidenced by participant feedback (\S~\ref{admin_sp_incidents}), some administrators have firsthand experience with data breaches, such as third-party integrations scraping user data (A4) or faculty auto-forwarding emails outside managed environments, thereby bypassing institutional phishing controls (A12). 
These incidents underscore the broader institutional vulnerabilities posed by unsanctioned apps, particularly regarding the leakage of personally identifiable information (PII) and violations of FERPA and other legal requirements.  

Interestingly, while many administrators expressed frustration with educators' use of unsanctioned apps, others acknowledged their own role in facilitating such practices. 
Some administrators even supported the integration of unsanctioned apps, validating educators' claims that they are occasionally encouraged to use these tools. 
This duality can create gray areas in policy enforcement and addressing this issue will require a collaborative approach between educators and administrators, while maintaining a common goal toward security.
By involving both educators and administrators in developing guidelines and reviewing tools, institutions can create realistic and effective policies that balance security requirements with the practical needs of teaching, ensuring compliance while supporting classroom needs.

\subsection{Conclusion}
Our findings emphasize the urgent need for institutions to rethink their approach to unsanctioned app use. 
Educators' preference for these tools, despite associated risks, reflects a demand for functionality and engagement that institutional offerings often fail to meet. 
At the same time, administrators' experiences with security incidents reveal critical gaps in policy implementation and enforcement. 
Educational institutions must invest in creating educator-friendly, while secure alternatives and provide clear pathways for vetting and approving education tools. 
Moreover, fostering a culture of shared responsibility---where both educators and administrators collaborate to balance innovation with compliance---will be essential for addressing these challenges effectively.

\bibliographystyle{unsrt}
\bibliography{references, references-extra, references-1, references-extra2, references-rakib}  
\appendix
\section{Appendix}
\label{app:study1ques}
\label{app:study2ques}
The complete questionnaires for the educator and administrator surveys are available at the following link: \url{https://osf.io/d9c6a/?view_only=a062ad724be64a82ab6033c556fa271e}.

\end{document}